\documentclass[conference]{IEEEtran}
\usepackage[T1]{fontenc}
\usepackage[utf8]{inputenc}
\usepackage{amsmath,amssymb}
\usepackage{graphicx}
\usepackage{array}
\usepackage[hidelinks]{hyperref}
\usepackage{cite}
\usepackage{microtype}

\title{Agent-ZTA: A Chain-Aware Zero Trust Model for Bidirectional Security-State Propagation and Runtime Containment in Multi-Agent Delegation}
\author{\IEEEauthorblockN{Asadullah Abdullah Khan, Toqeer Ali Syed}
	\IEEEauthorblockA{Faculty of Computer and Information Systems\\
		Islamic University of Madinah\\
		Madinah, Saudi Arabia\\
		471014356@stu.iu.edu.sa \quad | \quad toqeer@iu.edu.sa}}

\begin{document}
	\maketitle
	
	\begin{abstract}
		Multi-agent systems increasingly delegate tasks and permissions across chains of autonomous agents that invoke protected tools. Although agent identities and delegated credentials may remain valid throughout a workflow, the security condition of any agent along the chain can deteriorate during runtime execution—either through upstream compromise or downstream ingestion of untrusted inputs. Authorization models based solely on the immediate caller and its static delegated scope fail to account for these mid-execution state changes. This paper develops Agent-ZTA, a chain-aware Zero Trust authorization and containment model that links runtime security-state changes across multi-agent delegation graphs. The model distinguishes an agent’s path-specific delegated authority bound from the effective permissions currently exercisable under trusted policy. It defines an architecture featuring independently enforced policy decisions, authenticated delegation provenance, a dynamic chain-state view, and explicit fail-safe handling of missing or stale authorization signals. Furthermore, Agent-ZTA defines bidirectional security-state propagation, allowing downstream security events to affect relevant upstream workflow states, and specifies adaptive branch-scoped containment through restricted permissions and a sandbox ("jail") execution mode where an isolation boundary is available. The deterministic decision procedure, five stated security properties, and worked authorization traces show how accepted security-state changes constrain affected operations while preserving original delegation bounds and separating unrelated workflow branches under the defined policies. Finally, the paper states its trust assumptions and the exact scope of its security properties.
	\end{abstract}
	

	\section{Introduction}\label{sec:introduction}
	
	Autonomous artificial intelligence (AI) agents can plan multi-step tasks, invoke external tools, access application resources, and cooperate with other agents. In a multi-agent workflow, one agent may delegate a subtask to another, which can in turn delegate work to additional agents. This introduces chains of operational authority: individual agents act not only on their own behalf but also under permissions derived from earlier principals. As agentic systems become more capable, their security depends on governing the identities, permissions, and actions of these interacting principals, rather than protecting language-model outputs alone \cite{otsuka2026identity,chhabra2026security,schroeder2026challenges}.
	
	Zero Trust Architecture (ZTA) provides a useful starting point. Its principles emphasize explicit verification, least-privilege access, and decisions based on available information about the requesting subject, resource, and operating context \cite{nist2020zta}. Research applying these principles to AI agents has examined agent-specific identities, delegation provenance, capability constraints, runtime authorization, and enforcement at tool boundaries \cite{huang2026identity,nizamuddin2026identities,choong2026capchain,mishra2026ztmcp}. Other studies investigate risk-adaptive decisions and controls that remain independent of a potentially manipulated language model \cite{bhushan2026risk,chen2026swarmgov,muruaga2026bounded,dantuluri2026delegation}. Agent-ZTA builds on these mechanisms by connecting delegation dependencies to changing runtime security states, with branch-specific authorization and containment rules.
	
	Consider a user who delegates a task to Agent~A, which delegates a limited subtask to Agent~B; Agent~B subsequently delegates to Agent~C, which invokes a protected tool. If Agent~A becomes compromised after delegation, Agents~B and~C may retain valid credentials although the upstream execution state has degraded. The reverse case also matters: Agent~C may ingest malicious content and report a security event that changes how Agent~B or Agent~A should handle the affected workflow. These cases reveal a delegation-chain security-state problem: valid credentials alone do not establish whether the current operation or its dependent execution branch should remain authorized.
	
	Prior research establishes substantial adjacent capabilities. Work on authorization propagation considers transitive delegation and temporal constraints \cite{tallam2026propagation}; capability-oriented approaches govern field-level sharing, bounded delegated actions, or externally enforced delegated permissions, respectively \cite{choong2026capchain,muruaga2026bounded,dantuluri2026delegation}; and risk-adaptive approaches can revise individual access decisions as conditions change \cite{bhushan2026risk,chen2026swarmgov}. Security studies also show that adversarial content and harmful effects can traverse agent roles and workflow boundaries, underscoring the limitations of exclusively local protections \cite{shahroz2025siege,arif2026conjunctive,chang2026retrieval}. Together, these results motivate a model that links security-state events, authenticated delegation dependencies, and operation-specific authorization decisions in both directions along a workflow.
	
	Agent-ZTA addresses a concrete authorization dependency: a security event at one agent changes the state considered for the affected delegation branch, including relevant ancestors and descendants, while trusted policy determines which operations remain exercisable. Branch-scoped restrictions allow affected requests to be denied or confined to reduced-privilege execution while unrelated paths retain their independent authorization context.
	
	This paper develops \emph{Agent-ZTA}, a chain-aware Zero Trust authorization and containment model for autonomous multi-agent AI systems. The model distinguishes a downstream agent's maximum legitimately delegated authority, denoted $D_i$, from the authority it can currently exercise, denoted $E_i(t)$. The basic safety constraint is
	\begin{equation}
		E_i(t)\subseteq D_i.
		\label{eq:authority-bound}
	\end{equation}
	Section~\ref{subsec:authority-sets} makes the dependence on the delegation path $\pi$ and request context $c$ explicit, writing $D_i^{\pi}$ and $E_i^{\pi}(t,c)$. Agent-ZTA re-evaluates $E_i(t)$ using security-state information associated with the requesting agent and policy-relevant delegation dependencies. Downstream degradation can also mark a parent's affected workflow context for reassessment without equating this with compromise of the parent's entire identity. The model couples these state transitions to deterministic authorization, branch-scoped restriction, and a reduced-privilege jail mode for selected operations. Each effective permission set remains bounded by $D_i$, and policy decisions are evaluated outside the language model. Sections~\ref{sec:security-model} and~\ref{sec:mechanism} specify the state and authorization rules.
	
	This paper makes the following technical contributions:
	\begin{enumerate}
		\item \textbf{Delegation-chain problem formulation.} Distinguish the validity of delegated credentials from the current authorization state of an affected branch when an upstream or downstream agent degrades.
		\item \textbf{Bidirectional state model.} Define propagation of security-state effects along authenticated delegation dependencies, with explicit branch relevance and isolation from unrelated workflow paths.
		\item \textbf{Authorization and containment rules.} Define a deterministic request decision procedure and restricted jail-mode permission set, then derive bounded-authority properties and worked authorization traces.
	\end{enumerate}
	
	Section~\ref{sec:related} reviews the closest foundations and competing approaches, and Section~\ref{sec:threat} defines the threat model. Section~\ref{sec:security-model} specifies the security model, Section~\ref{sec:architecture} presents the reference architecture, and Section~\ref{sec:mechanism} details the authorization mechanism. Section~\ref{sec:properties} analyzes intended security properties and illustrative scenarios; Section~\ref{sec:limitations} discusses limitations and future evaluation, followed by the conclusion in Section~\ref{sec:conclusion}.
	
	\section{Related Work}\label{sec:related}
	
	\subsection{Zero Trust, Agent Identity, and Runtime Enforcement}
	
	NIST SP~800-207 frames Zero Trust as a security approach in which trust is not conferred solely by network location and access is evaluated explicitly at relevant boundaries \cite{nist2020zta}. For autonomous agents, this creates a distinction between \emph{establishing who is acting} and \emph{deciding whether that actor should perform a particular operation now}. Agent-specific identity research explores fine-grained credentials, accountable non-human principals, lifecycle controls, and the relationship between identity standards and agent behavior \cite{huang2026identity,nizamuddin2026identities,otsuka2026identity}. These approaches can establish the identity or origin of an agent and support traceability; they do not alone imply that a currently valid downstream credential remains safe to exercise after an upstream security incident. 
	
	Research on runtime enforcement moves policy decisions outside free-form model generation. ZT-MCP considers Zero Trust controls for MCP-connected tool access \cite{mishra2026ztmcp}; action-centric models emphasize governing tool calls and other agent actions rather than treating successful authentication as sufficient \cite{mok2026action}. SwarmGov-ZT examines runtime governance with changing trust information in a multi-agent cybersecurity setting \cite{chen2026swarmgov}. In transaction authorization, the Intent-Bound Withdrawal Authorization Protocol (IWAP) separates valid workload identity from permission to execute a particular withdrawal by requiring independently checked intent, transaction effects, and current admission state \cite{syed2026iwap}. A framework for AI-assisted vulnerability disclosure likewise binds its actions to current evidence revisions, human approvals, policy, and revocation state rather than treating an earlier approval as indefinitely sufficient \cite{syed2026fedtrust}. Agent-ZTA adopts independent policy enforcement as an architectural foundation, not as its claimed invention. Its additional concern is the source and interpretation of the security information used in a downstream access decision: relevant state may originate at an upstream delegator rather than the current tool-calling agent.
	
	\subsection{Delegation Chains, Capability Attenuation, and Revocation}
	
	Delegation is necessary when agents act on behalf of users or other agents. Foundational agent-delegation discussions emphasize authenticated delegation and the ability to distinguish the current actor from the principal on whose behalf it acts \cite{south2025delegation}. CapChain uses field-scoped capability tokens, verifiable provenance, and a capability-aware state reducer to constrain shared-state field access and forwarding between agents \cite{choong2026capchain}. It addresses a different access-control surface from downstream tool-authorization decisions. Work on bounded agents addresses delegation limits and the risks created by compositions of individually permissible actions \cite{muruaga2026bounded}. Delegation Without Trust studies sender-bound, attenuated credentials and external enforcement in the presence of potentially untrusted agent behavior \cite{dantuluri2026delegation}. These contributions are especially close to Agent-ZTA because they already address multi-hop authority, containment, and the need not to trust a language model with policy decisions.
	
	The critical distinction is between \emph{the bound of authority conferred by delegation} and \emph{the portion of that authority currently exercisable}. Capability attenuation can ensure that a child is not granted a broader scope than its parent. Credential expiration or revocation can terminate an authorization grant. Agent-ZTA does not replace these controls. Rather, it considers the policy case in which a downstream agent's credential has \emph{not} expired or been revoked, yet the state of a relevant upstream principal has deteriorated. A chain-aware mechanism must specify whether this change should restrict the downstream agent's effective authority before, or independently of, ordinary credential revocation. Whether a particular existing system already implements such a rule must be assessed from its actual policy semantics, not inferred merely from its use of delegation chains.
	
	Authorization Propagation in Multi-Agent AI Systems examines transitive relationships and changes in authorization validity at a workflow level \cite{tallam2026propagation}. Agent-ZTA extends the policy question beyond credential validity and transitive authorization: it binds accepted security events at either end of a delegation dependency to the affected workflow context. Upstream events can constrain downstream requests, while downstream events can trigger reassessment of dependent parent-workflow operations. The impact remains branch-scoped and operation-specific rather than automatically changing every principal's global security state.
	
	\subsection{Dynamic, Risk-Adaptive, and Continuous Authorization}
	
	Static role or capability checks are not always sufficient during long-running agent workflows. Risk-adaptive mechanisms can incorporate contextual information and changing risk indicators when deciding whether to allow an operation. Bhushan et al. describe risk-adaptive authorization for agentic AI systems, combining delegated access with policy-based decisions \cite{bhushan2026risk}. SwarmGov-ZT considers dynamically updated trust and governance information \cite{chen2026swarmgov}. Related preprint work explores uncertainty-aware decision-making and context-sensitive authorization \cite{fleming2025uncertainty}.
	
	These systems establish that \emph{continuous} or \emph{risk-aware} authorization alone cannot be claimed as novel. They also introduce an important comparison challenge: a sufficiently expressive generic policy engine could, in principle, consume upstream chain-state attributes if those attributes were supplied. Agent-ZTA's proposed contribution must therefore lie in an explicit model of which upstream agents are relevant, how changes are associated with delegation dependencies, and how resulting policies constrain effective downstream authority. Simply adding a numerical ``chain risk score'' to a conventional access request would not, by itself, establish a distinct mechanism.
	
	\subsection{Cross-Agent Attacks, Workflow Context, and Semantic Integrity}
	
	Security of an individual request does not imply security of the larger workflow. A broad examination of multi-agent security challenges identifies risks arising from interactions among otherwise legitimate components \cite{schroeder2026challenges}. The AgentDojo benchmark provides an environment for evaluating prompt-injection attacks against tool-using agents \cite{debenedetti2024agentdojo}. Other studies investigate attacks that exploit dependencies and interactions across agent tasks. Agents Under Siege studies optimized prompt attacks against multi-agent systems \cite{shahroz2025siege}, while Conjunctive Prompt Attacks investigates attacks whose effect depends on combining inputs across an agent system \cite{arif2026conjunctive}. Research on indirect prompt injection in deployed-style workflows further illustrates how untrusted content can affect agent actions \cite{chang2026retrieval}.
	
	These studies demonstrate that adversarial influence may cross component or agent boundaries. Their primary research object, however, is often attack success, prompt manipulation, or task integrity rather than the formal status of downstream delegated permissions. Similarly, semantic-state governance research considers situations in which policy checks and tool calls remain locally valid despite corrupted operational meaning \cite{doyle2026semantic}. Such work is adjacent to, but distinct from, the question of whether an upstream agent's degraded security condition should reduce a downstream delegate's exercisable authority. Agent-ZTA treats security-state information as an input to authorization; it does not propose a new prompt-injection detector or claim to solve semantic integrity generally.
	
	Defense papers also suggest useful design boundaries. Task Shield seeks to enforce task alignment against indirect prompt injection \cite{jia2025taskshield}, and IPIGuard applies tool-dependency information to constrain agent operations \cite{an2025ipiguard}. These mechanisms inform the need for explicit enforcement and workflow context. They should not be conflated with a delegated-authority mechanism unless their policies specifically track upstream principal security conditions and use them to alter downstream authorization.
	
	\subsection{Research Positioning and Design Focus}
	
	Existing research provides identity governance, authenticated multi-hop delegation, capability attenuation, revocation, request-level risk adaptation, and independent enforcement. Agent-ZTA combines these foundations with an explicit relation between delegation-path dependencies and security-state events at any participating agent. It specifies how affected upstream and downstream workflow contexts are re-evaluated and how sensitive operations can enter a restricted containment mode while unrelated branches remain separately authorized. This focus complements the multi-hop and bounded-delegation techniques of \cite{tallam2026propagation,muruaga2026bounded,dantuluri2026delegation,bhushan2026risk}.
	
	The central technical problem is to convert an accepted security event into a deterministic, operation-specific authorization restriction over the affected delegation branch. Agent-ZTA handles this by binding each request to its authenticated path, evaluating the security states relevant to that path, and intersecting the resulting restriction sets with its original delegated authority. Bidirectional state effects and branch-local containment distinguish this formulation from a purely caller-local authorization rule.
	
	\section{Threat Model}\label{sec:threat}
	
	\subsection{System Setting and Protected Assets}
	
	We consider an autonomous multi-agent workflow in which a user or service principal delegates a task to an agent, which may delegate narrower subtasks to further agents. Let a representative delegation path be $U\rightarrow A\rightarrow B\rightarrow C$, where $C$ requests an operation from a protected tool or resource. Each delegation conveys a bounded set of permissions. The authorization system can identify the requesting agent and associate its request with the relevant delegation path. A credential may remain valid throughout the scenario; validity alone does not imply that executing the requested operation remains appropriate after the chain's security condition changes.
	
	The principal asset to protect is the \emph{legitimate exercise of delegated authority} at tool or resource boundaries. This includes preserving original permission limits; re-evaluating dependent operations when security events originate upstream or downstream; containing affected workflow branches without automatically restricting unrelated branches; and maintaining an auditable basis for authorization decisions. Confidentiality, integrity, and availability of the underlying resources may be affected by improper tool use, but this paper does not claim to provide a comprehensive solution for every such failure.
	
	\subsection{Adversary Capabilities and Security-State Events}
	
	The adversary may influence or compromise an agent or its operating context at any position in a delegation path. Relevant examples include malicious instructions injected through an untrusted tool response, manipulation of retrieved or shared context, or an agent process operating under unauthorized control. These examples generate security-state events for the affected agent and the workflow dependencies associated with that event. The adversary may cause the affected agent to propose unauthorized subtasks, initiate misleading delegation requests, or attempt to induce downstream agents to misuse permissions already issued to them.
	
	One core case has an upstream agent degrade while downstream agents retain valid identities and delegated credentials. A second case has a downstream agent ingest a malicious input, triggering reassessment of the affected upstream workflow context. The adversary need not forge those credentials or obtain new permissions: the potential failure occurs when an authorization system treats the continuing validity of downstream authority as sufficient despite a security-relevant upstream change. We also allow a compromised upstream agent to issue additional requests \emph{within} the bounds it already holds. Ordinary out-of-scope actions and invalid credentials should be rejected by conventional enforcement and are not the distinguishing cases studied here.
	
	A \emph{security-state event} is an externally supplied indication that an agent's security condition has changed, such as a compromise flag, a policy-relevant anomaly, or an integrity-degradation signal. The authorization model consumes accepted events and propagates policy-relevant effects over authenticated delegation dependencies; detection remains a separate function. An event does not automatically imply that every downstream operation is malicious. Its authorization consequence depends on a later-defined deterministic policy and on the delegation relationship.
	
	\subsection{Trust Boundaries and Assumptions}
	
	The model uses the following trust boundaries and assumptions:
	\begin{enumerate}
		\item \textbf{Reliable identity and delegation provenance.} The system can identify the current requesting agent and retrieve an integrity-protected record of the relevant parent--child delegation relationships and original permission bounds. The paper does not propose a new cryptographic identity or credential format.
		\item \textbf{Trusted authorization infrastructure.} The policy decision and enforcement components, along with the source of authoritative policy and delegation records, are treated as trusted for the purposes of this model. An adversary controlling those components is outside scope.
		\item \textbf{Mediated resource access.} Protected operations pass through an enforcement point that can deny or constrain access according to the current decision. Direct, unmediated access to a tool is outside the guarantees contemplated by the model.
		\item \textbf{Available security-state inputs.} Policy-relevant state information and its association with agent identities can be supplied to the authorization system. The correctness, timeliness, and completeness of these inputs are assumptions or limitations, not established results.
		\item \textbf{Observable decision boundary.} The system can re-evaluate authorization at a controlled request or tool-invocation boundary. The proposal does not assume that previously completed actions can be reversed or that distributed actions can be interrupted instantaneously.
	\end{enumerate}
	
	Section~\ref{subsec:mech-freshness} defines the handling of missing, stale, or unverifiable required state inputs. The default is to deny the affected request, rather than infer a normal state from missing data. The policy determines which delegation dependencies and operations are affected; neither an upstream warning nor a downstream event automatically revokes the authority of every connected agent.
	
	\subsection{Illustrative Threat Scenario}
	
	Consider $U\rightarrow A\rightarrow B\rightarrow C\rightarrow T$, where $T$ is a protected tool. At time $t_0$, valid delegations authorize $C$ to invoke a limited operation on $T$. At a later time $t_1$, a relevant security-state signal indicates that $A$ is compromised or operating under degraded integrity. Neither $B$ nor $C$ has necessarily changed its local credentials, and $C$'s original delegated permissions remain valid.
	
	A baseline that checks only $C$'s identity, its delegated scope, and the current tool request could allow the invocation. That outcome would not mean the baseline is intrinsically insecure: it would indicate that the baseline has not been supplied with or configured to act on the upstream state change. Agent-ZTA evaluates whether $A$ is a policy-relevant ancestor of $C$ and whether the accepted event changes $C$'s \emph{currently exercisable} authority for this operation. The applicable restrictions are defined in Sections~\ref{subsec:bidirectional} and~\ref{sec:mechanism}; depending on the branch and requested action, the decision can remain unchanged or become more restrictive.
	
	This scenario isolates the targeted failure mode: \emph{valid downstream authorization artifacts coexist with a newly degraded upstream security condition}. It differs from a stolen credential, an attempt to exceed the original delegation bound, or an invalidly signed request, all of which can be addressed through existing identity and delegation checks.
	
	\subsection{Security Objectives and Out-of-Scope Threats}
	
	Within these assumptions, the authorization model preserves original delegation bounds, propagates relevant event effects across authenticated upstream and downstream workflow dependencies, selects operation-specific restrictions, confines affected branches to reduced-privilege execution when required, and enforces decisions outside agent-generated text.
	
	Outside scope are creating an intrusion-detection system, guaranteeing correct classification of compromises, repairing a manipulated agent's reasoning, preventing every prompt-injection attempt, securing resources that bypass the enforcement point, defending against compromise of the trusted policy infrastructure, and reversing operations already completed before a state change is observed. The effects of delayed or inaccurate security signals and the runtime cost of chain-aware decisions remain open questions outside the scope of this paper. These boundaries keep the mechanism focused on authorization and branch-specific containment.
	
	\section{Security Model}\label{sec:security-model}
	
	This section formalizes delegated authority, time-varying security states, bidirectional event effects, and branch-scoped containment. Sections~\ref{sec:architecture} and~\ref{sec:mechanism} connect these definitions to the authorization components and decision procedure.
	
	\subsection{Delegation Graph and Provenance}\label{subsec:delegation-graph}
	
	Let $V$ be a set of principals, including users, services, and agents, and let $G=(V,\mathcal{L})$ be a directed delegation graph. An edge $(u,v)\in\mathcal{L}$ records that principal $u$ has delegated some authority to principal $v$. Each edge is associated with an authorized scope $C_{u,v}\subseteq\mathcal{A}$, where $\mathcal{A}$ is the universe of protected operations (for example, resource--action pairs). Delegation records are assumed to have verifiable provenance and to satisfy any applicable validity conditions, as established in the threat model.
	
	A request $q=(i,a,c,t,\pi)$ identifies requesting agent $i\in V$, requested operation $a\in\mathcal{A}$, relevant request context $c$, decision time $t$, and the authenticated delegation path $\pi=(v_0,v_1,\ldots,v_k=i)$ relied upon for that request. When an agent has multiple delegation paths, $\pi$ identifies the \emph{specific path supporting this request}; unrelated paths are not silently combined to gain permissions. Which paths may legitimately support a request is determined by the underlying delegation policy, not by the agent's own assertion. The workflow branch $\gamma$ used by the containment rules is determined by $\pi$ and $c$ (Section~\ref{subsec:arch-flow}), so it is not a separate request field.

	For the chosen path, define the upstream and downstream principals of any $v_j$ on $\pi$; in particular, $\operatorname{Anc}_{\pi}(i)=\{v_0,\ldots,v_{k-1}\}$ for the requesting agent $i=v_k$:
	\begin{equation}
		\operatorname{Anc}_{\pi}(v_j)=\{v_0,\ldots,v_{j-1}\},\qquad
		\operatorname{Desc}_{\pi}(v_j)=\{v_{j+1},\ldots,v_k\}.
		\label{eq:ancestors}
	\end{equation}
	This path-specific definition matters: an incident affecting an agent outside $\pi$ is not automatically an incident affecting $i$'s authority under $\pi$.
	
	\subsection{Delegated Authority and Effective Authority}\label{subsec:authority-sets}
	
	The \emph{delegated authority bound} $D_i^{\pi}\subseteq\mathcal{A}$ is the maximum set of operations that agent $i$ may legitimately perform through path $\pi$, before applying current security-state restrictions. For a valid path, it is bounded by every edge scope along the path:
	\begin{equation}
		D_i^{\pi}\subseteq\bigcap_{j=1}^{k} C_{v_{j-1},v_j}.
		\label{eq:delegation-intersection}
	\end{equation}
	In particular, when $v_j$ delegates to $v_{j+1}$, the latter cannot acquire additional authority through that path: $D_{v_{j+1}}^{\pi_{j+1}}\subseteq D_{v_j}^{\pi_j}$, where $\pi_j$ denotes the corresponding prefix of $\pi$. These are established attenuation constraints, not the novel component of Agent-ZTA.
	
	The \emph{effective authority} $E_i^{\pi}(t,c)\subseteq\mathcal{A}$ is the set of operations that agent $i$ is authorized to exercise at decision time $t$, under context $c$ and path $\pi$. The central non-escalation invariant is
	\begin{equation}
		E_i^{\pi}(t,c)\subseteq D_i^{\pi}.
		\label{eq:effective-bound}
	\end{equation}
	An apparently valid credential establishes an admissible delegation bound; it does not, by itself, establish that the whole bound is exercisable at every later time.
	
	\subsection{Security-State Inputs and Chain Relevance}\label{subsec:state-inputs}
	
	Let $S_v(t)$ denote the security-state information available to the authorization system for principal $v$ at time $t$. The state vocabulary $\{\mathrm{normal},\mathrm{degraded},\mathrm{compromised},\mathrm{unavailable}\}$ records authorization-relevant conditions; the label \emph{unavailable} is assigned when a required record is absent, unverifiable, or outside its freshness window (Section~\ref{subsec:mech-freshness}) and is never inferred as normal. Each accepted event is bound to a principal, affected delegation-path or workflow context, observation time, and provenance. State-event acceptance is distinct from intrusion detection.
	
	Not every observed event affects every operation or delegation path. Define $R_{\pi}(i,a,c,t)\subseteq\operatorname{Anc}_{\pi}(i)$ as the upstream principals whose reported states are policy-relevant to operation $a$ for agent $i$ under path $\pi$ and context $c$. The resulting relevant chain-state input is
	\begin{equation}
		\Sigma_{i,\pi}(a,c,t)=\bigl\{(v,S_v(t)):v\in R_{\pi}(i,a,c,t)\bigr\}.
		\label{eq:relevant-states}
	\end{equation}
	The relevance relation is controlled by trusted policy. For the same chain, an ancestor's degraded condition can restrict a sensitive write while leaving an unrelated read unaffected. Accepted downstream events can additionally mark a specific upstream workflow context for reassessment; this contextual taint does not equate to compromise of the parent principal across all its tasks.
	
	\subsection{Bidirectional Event Propagation and Branch State}\label{subsec:bidirectional}
	
	Let an accepted event be $e=(v,\gamma,s,\tau,\rho)$, where $v$ identifies the affected principal, $\gamma$ identifies the authenticated workflow or delegation context, $s$ is the reported condition, $\tau$ is the observation time, and $\rho$ is its provenance. For a path $\pi$ and a principal $v$ on it, $\operatorname{Anc}_{\pi}(v)$ and $\operatorname{Desc}_{\pi}(v)$ are its upstream and downstream principals on $\pi$ (Eq.~\eqref{eq:ancestors}). Trusted policy selects the event's affected context and relevant dependencies using
	\begin{equation}
		\mathcal{I}_P(e,\pi,a,c)\subseteq
		\{v\}\cup\operatorname{Anc}_{\pi}(v)\cup\operatorname{Desc}_{\pi}(v).
		\label{eq:influence}
	\end{equation}
	The set $\mathcal{I}_P$ represents \emph{authorization impact}, not an assertion that every included principal is independently compromised. An event at an upstream delegator can restrict descendants relying on that delegation. Conversely, an event at a downstream agent can taint the parent workflow context associated with the affected branch, causing its parent to re-evaluate actions that depend on that branch. An unrelated branch without the affected dependency is not included automatically.
	
	For each authenticated workflow context $\gamma$, define an accepted branch-condition record $T_{\gamma}(t)$ assembled from relevant accepted events. The PDP reads $T_{\gamma}(t)$ together with principal states when applying policy. Relevance, event provenance, freshness, and branch association are verified before this record influences an authorization decision. Reaching an ancestor through $\mathcal{I}_P$ does not overwrite that ancestor's identity or security state in unrelated workflow contexts.
	
	\subsection{Branch-Scoped Containment State}\label{subsec:jail-model}
	
	Let $M_{\gamma}(t)\in\{\mathrm{normal},\mathrm{restricted},\mathrm{jail}\}$ be the containment mode of the affected workflow branch, assigned by trusted policy from the accepted branch record:
	\begin{equation}
		M_{\gamma}(t)=\mu_P\bigl(T_{\gamma}(t)\bigr),
		\label{eq:mode}
	\end{equation}
	where $\mu_P$ maps a branch record to a mode and returns $\mathrm{normal}$ when no relevant event has been accepted for $\gamma$. Let $J_P(\gamma,c)\subseteq\mathcal{A}$ be the operations allowed in jail mode; the policy excludes sensitive operations from this set and may choose $J_P=\varnothing$. Define
	\begin{equation}
		C_P(\gamma,c,t)=\begin{cases}
			\mathcal{A}, &M_{\gamma}(t)=\mathrm{normal},\\
			Q_P(\gamma,c), &M_{\gamma}(t)=\mathrm{restricted},\\
			J_P(\gamma,c), &M_{\gamma}(t)=\mathrm{jail},
		\end{cases}
		\label{eq:branch-containment}
	\end{equation}
	where $Q_P(\gamma,c)\subseteq\mathcal{A}$ is a configured restricted-mode allowance. Because $M_{\gamma}$ is determined by $T_{\gamma}$, event effects propagated in either direction reach the authorization rule through $C_P(\gamma,c,t)$. The resulting permission set is always intersected with the legitimate delegation bound. A jail mode is therefore an authorization restriction; execution isolation additionally requires an enforcement environment that prevents access outside the restricted boundary. The model does not infer such isolation solely from a change in policy label.
	
	\subsection{Chain-Aware Authorization Function}\label{subsec:decision-function}
	
	Let $P$ denote the trusted authorization policy. Formally, $P$ computes a set of currently permitted operations $K_P(i,\pi,c,t,\mathbf{S})\subseteq\mathcal{A}$ from the requesting agent's own security information, the policy-relevant upstream information, and other applicable context. We define effective authority by intersection, incorporating the branch-mode restriction:
	\begin{equation}
		\begin{aligned}
			E_i^{\pi}(t,c)&=D_i^{\pi}\cap C_P(\gamma,c,t)\\
			&\quad\cap K_P\bigl(i,\pi,c,t,S_i(t),\\
			&\qquad\Sigma_{i,\pi}(\cdot,c,t)\bigr).
		\end{aligned}
		\label{eq:effective-policy}
	\end{equation}
	The dot indicates that relevance and the allowed-set calculation may depend on the operation being considered; the branch context $\gamma$ is authenticated with the request. Equivalently, a specific request is permitted only if its operation is within the legitimate delegation bound \emph{and} its current chain-aware policy check passes:
	\begin{equation}
		\begin{aligned}
			\operatorname{Permit}(q)\iff {}&a\in D_i^{\pi}\land a\in C_P(\gamma,c,t)\land\\
			&a\in K_P\bigl(i,\pi,c,t,S_i(t),\\
			&\qquad\Sigma_{i,\pi}(\cdot,c,t)\bigr).
		\end{aligned}
		\label{eq:permit}
	\end{equation}
	Here $P$ is deterministic for a fixed, authoritative input snapshot. Section~\ref{subsec:mech-freshness} specifies how the PDP selects accepted input snapshots and checks required provenance and freshness; no instantaneous global state consistency is assumed.
	
	This formulation exposes the intended distinction from a purely local baseline. The baseline may evaluate a local policy $P_{\mathrm{local}}$ on $(i,a,c,t,S_i(t))$ only, whereas Agent-ZTA additionally binds the decision to $\pi$, the relevant upstream states, and the branch allowance $C_P(\gamma,c,t)$, which is determined by the accepted branch record $T_{\gamma}(t)$. The distinction concerns the \emph{decision inputs and the treatment of delegation dependencies}, not a claim that existing policy engines lack conditional-access capabilities.
	
	\subsection{State Changes, Restriction, and Reassessment}\label{subsec:state-transition}
	
	Let $t_0<t_1$ and suppose an upstream principal $u\in\operatorname{Anc}_{\pi}(i)$ has a reported security-state change between those times. If $u$ is relevant to a requested operation, the policy can reassess the operation using the new state information even when $D_i^{\pi}$, $i$'s identity, and its credentials have not changed. The resulting authority may be unchanged, reduced for some operations, or empty (suspension or denial):
	\begin{equation}
		E_i^{\pi}(t_1,c)\in\mathcal{P}(D_i^{\pi}),
		\label{eq:state-outcome}
	\end{equation}
	where $\mathcal{P}(D_i^{\pi})$ denotes the set of subsets of the original delegation bound. In particular, $E_i^{\pi}(t_1,c)=\varnothing$ represents no currently exercisable operations under that path and context; it need not imply permanent deletion of a credential.
	
	For a \emph{fixed policy, delegation path, and other decision inputs}, a policy may be designed so that security degradation cannot increase the authority it permits. This optional monotonicity condition can be written as
	\begin{equation}
		S'\preceq_{\mathrm{trust}}S\quad\Longrightarrow\quad
		K_P(S')\subseteq K_P(S),
		\label{eq:degradation-monotonicity}
	\end{equation}
	where $S'\preceq_{\mathrm{trust}}S$ means that $S'$ represents no more trustworthy security conditions than $S$. This is a \emph{proposed policy property under stated assumptions}, not a claim that all contexts, policies, and state labels admit a total trust ordering. Recovery or a newly authorized context may legitimately change effective permissions, but cannot expand them beyond $D_i^{\pi}$.
	
	Propagation is evaluated in both directions across authenticated delegation dependencies: upstream events affect policy-relevant descendants, and downstream events taint the associated parent workflow context. Both directions update subsequent mediated decisions and can change $M_{\gamma}(t)$; neither rewrites the original delegated authority.
	
	\subsection{Worked Authorization Example and Model Boundaries}\label{subsec:model-example}
	
	Consider $U\rightarrow A\rightarrow B\rightarrow C$, with a valid path-specific bound $D_C^{\pi}=\{\mathrm{read},\mathrm{write}\}$ on a protected resource. At $t_0$, assume the relevant policy allows both operations. At $t_1$, an externally supplied state signal indicates that $A$ is degraded. For an \emph{illustrative} policy in which $A$ is relevant to write authorization but not to read authorization, the model may yield
	\begin{align}
		E_C^{\pi}(t_0,c)&=\{\mathrm{read},\mathrm{write}\},\\
		E_C^{\pi}(t_1,c)&=\{\mathrm{read}\}.
		\label{eq:example-authority}
	\end{align}
	The credentials and bound $D_C^{\pi}$ remain unchanged. The new decision arises solely from the upstream state input and applicable policy. A stricter policy might instead yield $E_C^{\pi}(t_1,c)=\varnothing$; a policy that considers $A$ irrelevant to the requested operation might leave the decision unchanged. These are worked policy cases derived from the stated predicates, not measurements.
	
	The definitions establish the state, path, and branch-mode constraints used by the decision procedure in Section~\ref{sec:mechanism}. Event production, trust in input provenance, and state freshness remain separate operational dependencies.

	\section{Agent-ZTA Architecture}\label{sec:architecture}
	
	This section defines the Agent-ZTA authorization and containment architecture. The design integrates authenticated delegation paths, bidirectional branch-state updates, independent policy evaluation, and restricted execution modes. The security model in Section~\ref{sec:security-model} supplies its decision inputs and invariants.
	
	\subsection{Architectural Overview}\label{subsec:arch-overview}
	
\begin{figure*}[!t]
		\centering
		\includegraphics[width=0.96\textwidth]{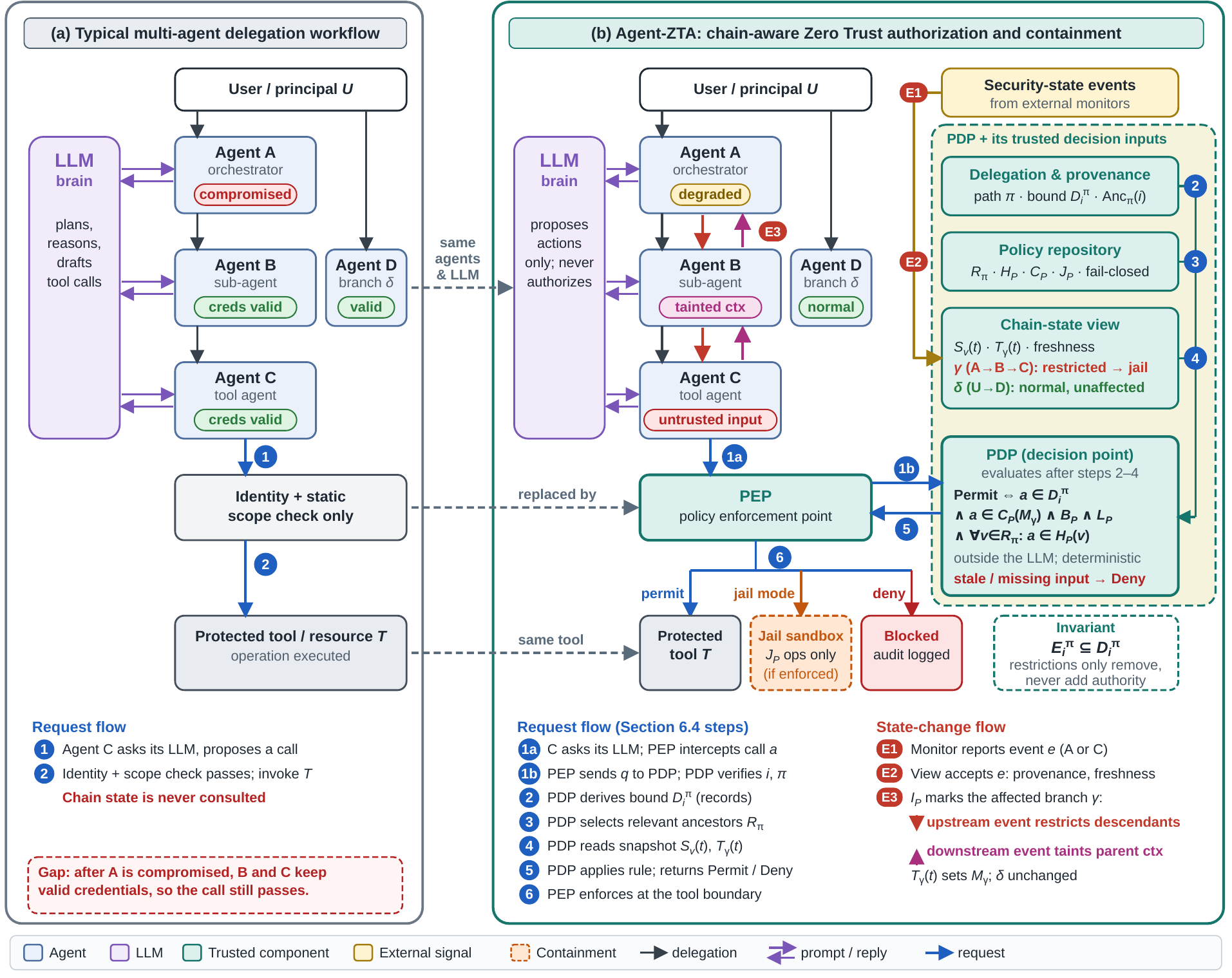}
		\caption{Agent-ZTA architecture for the delegation chain $U\rightarrow A\rightarrow B\rightarrow C$ with an unrelated branch through $D$. (a) A typical multi-agent workflow: agents consult an LLM, and the tool is reached after an identity and static-scope check only. (b) Agent-ZTA: the LLM only proposes actions, while a PEP and a deterministic PDP decide using delegation provenance, trusted policy, and the chain-state view. Request flow, mapped to the steps of Section~\ref{subsec:mech-procedure}: 1a--1b authenticate and bind (the PEP intercepts $C$'s call and sends $q$ to the PDP); 2 check the delegated bound $D_i^{\pi}$; 3 select relevant upstream principals $R_{\pi}$; 4 read the decision snapshot ($S_v$, $T_{\gamma}$); 5 apply branch containment and restrictions and return Permit or Deny; 6 enforce at the tool boundary (permit, jail mode, or deny). State-change flow: E1 an external monitor reports an event; E2 the chain-state view accepts it after provenance and freshness checks; E3 $\mathcal{I}_P$ marks the affected branch $\gamma$ (red: an upstream event restricts descendants; pink: a downstream event taints the parent context) and $T_{\gamma}$ sets $M_{\gamma}$ (Sections~\ref{subsec:bidirectional} and~\ref{subsec:jail-model}). Dashed connectors link components that are the same, or replaced, across the two panels.}
		\label{fig:architecture}
	\end{figure*}

	Figure~\ref{fig:architecture} contrasts two designs for the same workflow. In panel (a), a typical multi-agent delegation workflow, each agent consults a language model to plan and draft tool calls, and a protected tool is reached after only an identity and static-scope check. Panel (b) shows the Agent-ZTA authorization boundary for a protected tool invocation by a downstream agent. Agents plan and delegate tasks, but neither language-model output nor self-reported security status is authoritative for access control. A policy enforcement point (PEP) mediates protected operations; a policy decision point (PDP) evaluates authenticated delegation provenance, applicable permission bounds, accepted principal and branch-state records, and trusted policy. The same decision boundary applies when an accepted event arises upstream or downstream: the event affects only the policy-relevant workflow context and operations. The request path (steps 1a--6 in the figure) follows the decision procedure of Section~\ref{subsec:mech-procedure}, and the separate state-change path (E1--E3) updates the chain-state view. The supporting records are distinct responsibilities, not necessarily separate storage services.
	
	\subsection{Components and Responsibilities}\label{subsec:arch-components}
	
	The architecture separates six principal responsibilities:
	\begin{enumerate}
		\item \textbf{Agent workflow and protected tools.} Agents execute delegated tasks and request operations on protected resources. Each protected operation must pass through an authorization boundary; the architecture does not assume that agents can be trusted to police their own tool calls.
		\item \textbf{Policy enforcement point (PEP).} The PEP intercepts a tool or resource request, supplies its authenticated request information for evaluation, and executes the operation only when the returned authorization decision permits it. A denial does not depend on an LLM obeying a natural-language warning.
		\item \textbf{Policy decision point (PDP).} The PDP evaluates whether the requested operation lies within the authenticated path-specific bound $D_i^{\pi}$ and satisfies the chain-aware policy $P$ of Eq.~\eqref{eq:permit}. The PDP also applies the accepted branch condition and selected containment mode.
		\item \textbf{Delegation and provenance records.} Trusted records identify the delegation edges, permitted scopes, and path $\pi$ that legitimately supports the request. They provide the basis for deriving $D_i^{\pi}$ and $\operatorname{Anc}_{\pi}(i)$. A requesting agent must not be allowed to invent or alter its own upstream path to avoid a relevant restriction.
		\item \textbf{Chain-state view.} This component associates accepted principal-state events $S_v(t)$, branch-state records $T_{\gamma}(t)$, delegation contexts, provenance, and freshness metadata. It supports event effects in either direction across the authenticated chain.
		\item \textbf{Policy repository.} Trusted policy defines which ancestors are relevant to which operations, how their reported states influence the decision, and how missing or outdated information is treated. Policy also defines affected-branch selection and restricted or jail-mode allowances, without imposing uniform revocation on unrelated paths.
	\end{enumerate}
	
	Identity validation, credential verification, and protected audit records may be provided by existing infrastructure. They are necessary supporting controls, but the paper does not introduce new cryptographic protocols, independent monitoring systems, or a new language-model security classifier.
	
	\subsection{Authorization and State-Change Paths}\label{subsec:arch-flow}
	
	At a protected request boundary, the PEP receives an operation $a$, requesting agent $i$, and context $c$. The authorization infrastructure verifies the requester and the delegation path $\pi$, derives the legitimate bound $D_i^{\pi}$, and associates the request with its authenticated workflow branch $\gamma$. The PDP reads the accepted security-state snapshot, including the requester's condition, policy-relevant ancestors, and $T_{\gamma}(t)$, which captures accepted bidirectional event effects. It then applies the branch allowance $C_P(\gamma,c,t)$ and the authorization rule in Eq.~\eqref{eq:mechanism-permit}. The PEP permits the protected operation only when the resulting decision is \textsc{Permit}.
	
	A security event follows a separate state-input path. An upstream event at $A$ can constrain a request by $C$ when $A\in R_{\pi}(i,a,c,t)$; a downstream event at $C$ can mark the dependent parent-workflow context for reassessment through $\mathcal{I}_P$ in Eq.~\eqref{eq:influence}. Both cases update the accepted branch view rather than changing existing delegation credentials. Under a selected policy, a degraded branch may retain a read operation while denying a sensitive write, or enter a restricted or jail mode. These are outcomes of the specified policy predicates for the accepted state, not empirical measurements.
	
	A security-state update affects an authorization decision when the PDP has accepted that update at a mediated request boundary. The snapshot, freshness, and required-input rules are specified in Section~\ref{subsec:mech-freshness}. The model does not assume instantaneous delivery to all dependent branches or cancellation of operations already in progress.
	
	\subsection{Design Invariants and Trust Boundaries}\label{subsec:arch-invariants}
	
	Four constraints guide the architecture. \emph{Delegation-bounded authorization} requires $E_i^{\pi}(t,c)\subseteq D_i^{\pi}$ independently of security-state changes. \emph{Bidirectional dependency relevance} binds accepted upstream or downstream events to the affected authenticated workflow context without treating every connected principal as compromised. \emph{Branch-scoped containment} intersects effective authority with $C_P(\gamma,c,t)$ while separately evaluating unrelated branches. \emph{Independent enforcement} requires a protected operation to be mediated by a trusted PEP acting on a deterministic PDP decision rather than on agent-generated instructions. These constraints are evaluated under the trusted-component assumptions in Section~\ref{sec:threat}.
	
	The trusted boundary comprises policy, delegation provenance, the decision and enforcement components, and the integrity of information provided to them. Agent-generated content and external inputs are not authoritative authorization sources. The quality and timing of external security-state signals remain assumptions: Agent-ZTA governs the \emph{authorization consequences} of accepted events from either direction, rather than detecting those events. Jail mode restricts authorized operations; physical process or network isolation additionally requires an enforcing sandbox boundary.
	
	\subsection{Scope of the Proposed Architecture}\label{subsec:arch-scope}
	
	The principal distinction from caller-local authorization is the combined use of an authenticated delegation dependency $\pi$, operation-relevant upstream state $\Sigma_{i,\pi}(a,c,t)$, and accepted branch state $T_{\gamma}(t)$ when deciding effective authority. The branch record captures relevant event impacts in both directions, while $C_P(\gamma,c,t)$ restricts selected operations in normal, restricted, or jail mode. The model reuses existing PEP/PDP and credential mechanisms; its contribution is the decision structure that associates state changes with branch-specific authority.
	
	Section~\ref{sec:mechanism} specifies the resulting decision semantics and state-handling rules. 
	
	\section{Chain-Aware Authorization Mechanism}\label{sec:mechanism}
	
	This section specifies the deterministic authorization and containment decision procedure that applies the security model in Section~\ref{sec:security-model}. Each request is bound to a verified delegation path and workflow context, the accepted bidirectional branch-state record, and a policy-selected containment mode.
	
	\subsection{Decision Inputs and Policy Responsibilities}\label{subsec:mech-inputs}
	
	For each protected request $q=(i,a,c,t,\pi)$, the PEP forwards authenticated request attributes to the PDP. The PDP obtains: (i) the verified path $\pi$ and path-specific delegated bound $D_i^{\pi}$ from trusted delegation records; (ii) the applicable policy version $P$; (iii) security-state records for the requesting agent and each upstream principal in $R_{\pi}(i,a,c,t)$; and (iv) the accepted branch-condition record $T_{\gamma}(t)$ for the authenticated branch $\gamma$, from which policy derives $M_{\gamma}(t)$. A state record includes an asserted condition, principal identifier, and available provenance and observation-time metadata. Signal production and compromise detection are outside Agent-ZTA's proposed contribution.
	
	The trusted policy identifies operation-relevant ancestors and affected workflow branches and then selects the permissible operations using the accepted security-state records and branch-containment allowance. In particular, an agent cannot select an alternative path merely to evade a restriction, and it cannot supply its own authoritative state or policy. Where multiple legitimate paths exist, each request is evaluated against the particular authenticated path on which the requested authority depends; this paper does not assume that permissions from independent paths can be combined.
	
	\subsection{Bounded Chain-State Restriction Rule}\label{subsec:mech-rule}
	
	The policy representation makes the decision rule explicit. Let $B_P(i,a,c)$ denote the ordinary local and contextual permission predicate, and let $L_P(i,c,S_i)$ be the set of operations allowed by the requester's own reported security condition. For each relevant ancestor $v$, let $H_P(v,i,\pi,c,S_v)$ be the set of operations the policy permits downstream of $v$ under that reported condition. These functions are \emph{policy-defined restrictions}, not new credentials or security detectors. Together they expand the generic policy set $K_P$ of Section~\ref{subsec:decision-function}: $K_P=\{a\in\mathcal A:\ B_P(i,a,c),\ a\in L_P(i,c,S_i),\ \forall v\in R_{\pi}(i,a,c,t):\ a\in H_P(v,i,\pi,c,S_v)\}$, so Eq.~\eqref{eq:mechanism-permit} is the single-operation form of Eq.~\eqref{eq:permit}. The accepted branch record $T_{\gamma}(t)$ enters through $C_P(\gamma,c,t)$ via $M_{\gamma}(t)=\mu_P(T_{\gamma}(t))$ (Eq.~\eqref{eq:mode}). For a fixed, accepted input snapshot $\widehat{\mathbf S}$, define the effective set \emph{operation by operation}:
	\begin{equation}
		\begin{aligned}
			\widehat E_i^{\pi}&=D_i^{\pi}\cap C_P(\gamma,c,t)\\
			&\quad\cap\bigl\{a\in\mathcal A:\ B_P(i,a,c),\\
			&\qquad a\in L_P(i,c,\widehat S_i),\\
			&\qquad \forall v\in R_{\pi}(i,a,c,t),\\
			&\qquad a\in H_P(v,i,\pi,c,\widehat S_v)\bigr\}.
		\end{aligned}
		\label{eq:mechanism-effective}
	\end{equation}
	The operation-specific relevance set $R_{\pi}(i,a,c,t)$ is the one defined in Section~\ref{subsec:state-inputs}, evaluated under the applicable policy and accepted snapshot. Defining the set operation by operation avoids imposing an ancestor's restrictions on a different operation for which that ancestor is not relevant.
	
	Intersection makes the non-escalation property immediate: $\widehat{E}_i^{\pi}\subseteq D_i^{\pi}$ regardless of the state values. In the illustrative policy family used here, a degraded state may remove certain operations from an ancestor's downstream allowance, while a compromised state may remove more. This is a \emph{choice of policy}, not a universal ordering of all possible security signals. A report of degradation does not automatically revoke every descendant or every operation.
	
	\subsection{State Selection, Freshness, and Missing Inputs}\label{subsec:mech-freshness}
	
	The PDP evaluates each request against a \emph{decision snapshot}: the verified path and delegation scopes, a specific policy version, and the security-state records accepted for that decision. For each required state record, the policy specifies whether its source is acceptable and whether its age is within a defined freshness condition. An absent, unverifiable, or out-of-policy-age required record is classified as \emph{unavailable}, rather than silently treated as normal. The policy must specify a conservative outcome for unavailable inputs; in the proposed default for a protected operation, a required unavailable input results in denial of that operation until sufficient information is available. This is a formal fail-closed rule, not a claim of immediate revocation or globally synchronized storage.
	
	State records can arrive asynchronously. Consequently, a completed decision reflects only the accepted records visible at its decision boundary. The mechanism does not guarantee that a newly detected compromise is known to all PDPs immediately, that a permit remains safe indefinitely, or that an already executed action can be reversed. Where a new security-state record becomes visible, subsequent affected requests are re-evaluated using the newer accepted snapshot. An implementation may use versions or expiration information to distinguish records, but no particular distributed synchronization protocol is prescribed here.
	
	\subsection{Per-Request Decision Procedure}\label{subsec:mech-procedure}
	
	The deterministic procedure proceeds as follows:
	\begin{enumerate}
		\item \textbf{Authenticate and bind.} The PEP intercepts the request and supplies the requesting principal, operation, and context. The PDP verifies the principal and the authenticated delegation path $\pi$. If identity or path verification fails, return \textsc{Deny}. (Fig.~\ref{fig:architecture}, steps 1a--1b.)
		\item \textbf{Check delegated bounds.} Derive $D_i^{\pi}$ from trusted, currently valid delegation records. If $a\notin D_i^{\pi}$, return \textsc{Deny} without attempting to enlarge the bound through a security-state policy. (Fig.~\ref{fig:architecture}, step 2.)
		\item \textbf{Select relevant upstream principals.} Under the applicable trusted policy, obtain $R_{\pi}(i,a,c,t)\subseteq\operatorname{Anc}_{\pi}(i)$. This set is computed from the verified path and request attributes, not from the requester's claimed security condition. (Fig.~\ref{fig:architecture}, step 3.)
		\item \textbf{Read a decision snapshot.} Obtain accepted principal records, relevant upstream conditions, and branch-state record $T_{\gamma}(t)$, including downstream-to-parent event effects. Reject missing, stale, or unverifiable required inputs. (Fig.~\ref{fig:architecture}, step 4.)
		\item \textbf{Apply branch containment and restrictions.} Determine $M_{\gamma}(t)=\mu_P(T_{\gamma}(t))$ and check $a\in C_P(\gamma,c,t)$, $B_P(i,a,c)$, $a\in L_P(i,c,\widehat S_i)$, and every relevant ancestor predicate $a\in H_P(v,i,\pi,c,\widehat S_v)$. Return \textsc{Permit} only if every condition holds; otherwise return \textsc{Deny}. (Fig.~\ref{fig:architecture}, step 5.)
		\item \textbf{Enforce at the boundary.} The trusted PEP executes the requested operation only on a \textsc{Permit} decision for that request and snapshot. A later request requires a new decision; a prior permit is not a permanent authorization grant. (Fig.~\ref{fig:architecture}, step 6.)
	\end{enumerate}
	
	For valid and available inputs, the single-operation rule is therefore
	\begin{equation}
		\begin{aligned}
			\operatorname{Permit}(q)\iff{}&a\in D_i^{\pi}\\
			&\land a\in C_P(\gamma,c,t)\\
			&\land B_P(i,a,c)\\
			&\land a\in L_P(i,c,\widehat S_i)\\
			&\land \bigwedge_{v\in R_{\pi}(i,a,c,t)}\\
			&\qquad[a\in H_P(v,i,\pi,c,\widehat S_v)].
		\end{aligned}
		\label{eq:mechanism-permit}
	\end{equation}
	The rule is evaluated only after the required identity, path, record-validity, and freshness checks succeed; failure of any required preliminary check yields \textsc{Deny}. The result is deterministic \emph{given the same verified inputs and policy version}, not necessarily identical across decisions made with different snapshots.
	
	\subsection{Upstream Changes, Recovery, and Scope of Propagation}\label{subsec:mech-changes}
	
	Suppose an accepted report changes the state of upstream agent $u$ from normal to degraded while $i$'s credentials and $D_i^{\pi}$ remain unchanged. If $u\notin R_{\pi}(i,a,c,t)$, the event does not directly alter the decision for $a$ through this mechanism. If $u\in R_{\pi}(i,a,c,t)$, the PDP applies $H_P$ to the new accepted state on the next affected decision. An operation may then be denied, or remain allowed if the policy does not restrict that operation under the new condition. This is \emph{reauthorization of descendants through a delegation dependency}; it is not credential revocation or an assertion that every descendant must be blocked.
	
	For a fixed path, policy, context, and all other state inputs, a useful security design condition is that an ancestor's degradation does not increase its downstream allowance: if $S'_u\preceq_{\mathrm{trust}} S_u$, then $H_P(u,i,\pi,c,S'_u)\subseteq H_P(u,i,\pi,c,S_u)$. In this restricted setting, Eq.~\eqref{eq:mechanism-effective} implies that the resulting effective authority cannot increase. This is an analytical consequence of \emph{the specified policy condition}, not a proof of end-to-end security under faulty signals or concurrent updates. Recovery may restore some operations after an accepted improvement in the relevant state; any restored authority still remains within $D_i^{\pi}$. No time-to-recovery or automatic trust-restoration guarantee is assumed.
	
	\subsection{Reverse Propagation and Restricted Branch Execution}\label{subsec:mech-reverse}
	
	For an accepted downstream event at agent $v$, the PDP determines the authenticated branch $\gamma$ on which the affected execution depends and updates its accepted event view. Under $\mathcal{I}_P$ in Eq.~\eqref{eq:influence}, an upstream delegator is marked for reassessment \emph{only for that dependent branch}. A parent with an unrelated task or delegation path retains a separate decision context. This approach models reverse propagation as a change in authorization-relevant workflow state, not an unsupported assertion that compromise of a child necessarily compromises its parent process.
	
	Trusted policy sets $M_{\gamma}(t)$ to $\mathrm{restricted}$ or $\mathrm{jail}$ when a required event condition is met. The PDP then intersects the request's allowance with $C_P(\gamma,c,t)$ in Eq.~\eqref{eq:branch-containment}. For example, a jailed branch may be permitted only to read approved diagnostic material while write, network publication, and new delegation are denied. If an execution sandbox is available and enforced, the branch is routed to that restricted environment; absent such an enforcement boundary, the algorithm guarantees only the stated authorization denial, not process isolation. Operations on unrelated branches are evaluated using their own authenticated branch contexts.
	
	\subsection{Illustrative Trace and Mechanism Boundaries}\label{subsec:mech-trace}
	
	Consider the verified path $U\rightarrow A\rightarrow B\rightarrow C$ and $D_C^{\pi}=\{\mathrm{read},\mathrm{write}\}$. Assume the policy makes $A$ relevant to a write request but not to a read request, and assume all required records are acceptable unless stated. At $t_0$, $A$ is reported normal and both operations pass the policy. At $t_1$, an accepted record reports $A$ as degraded; under the \emph{illustrative} $H_P$, write is excluded while read remains permitted, and neither $C$'s delegation bound nor its credential changes. At $t_2$, the record for $A$ that is required for write becomes unavailable (absent or stale), and the fail-closed rule denies write. A second case starts with a security event at $C$ after ingestion of an untrusted input. Policy marks $B$'s workflow context on this same branch for reassessment and sets $M_{\gamma}(t)=\mathrm{jail}$, so the branch loses write authority under $J_P$ while a separate unaffected workflow path remains subject to its own policy inputs. Table~\ref{tab:mechanism-trace} lists these cases together with two boundary cases: an unrelated degraded principal and an operation outside the original scope.
	
	\begin{table}[!t]
		\centering
		\caption{Worked authorization decisions for $U\rightarrow A\rightarrow B\rightarrow C$ under the specified policy predicates (not empirical measurements).}\label{tab:mechanism-trace}
		\footnotesize
		\begin{tabular}{@{}>{\raggedright\arraybackslash}p{0.27\columnwidth}>{\raggedright\arraybackslash}p{0.33\columnwidth}>{\raggedright\arraybackslash}p{0.30\columnwidth}@{}}
			\hline
			Situation & Relevant policy inputs & Outcome \\
			\hline
			All accepted states normal ($t_0$) & Valid path; $A$ normal & Read and write permitted \\
			Upstream $A$ degraded ($t_1$) & Accepted record; $A$ relevant to write only & Read permitted; write denied \\
			Required state unavailable ($t_2$) & Record for $A$ missing or stale for write & Write denied (fail-closed) \\
			Event at $C$ (untrusted input) & Accepted event; $M_{\gamma}=\mathrm{jail}$ & Write denied on branch $\gamma$; other branches unchanged \\
			Unrelated principal degraded & Principal not on the authenticated path & No restriction from that event \\
			Write outside original scope & $\mathrm{write}\notin D_C^{\pi}$ & Write denied regardless of state \\
			\hline
		\end{tabular}
	\end{table}
	
	The mechanism assumes trusted provenance and policy inputs, mediated tool access, and an accepted decision snapshot. These assumptions delimit when branch restriction and jail-mode permissions are enforceable. Independent signal generation and physical sandbox isolation require corresponding external controls.

	\section{Security Properties and Illustrative Scenarios}\label{sec:properties}
	
	This section states authorization properties and the assumptions under which they hold, using the state model and decision rule in Sections~\ref{sec:security-model} and~\ref{sec:mechanism}. The arguments assume trusted PEP/PDP components, authenticated delegation provenance, a fixed policy version, and an accepted state snapshot at the decision boundary.
	
	\subsection{Delegation-Bound Preservation}\label{subsec:prop-bound}
	
	\textbf{Property 1 (non-escalation within a path).} For any request context, policy, and accepted security-state snapshot, the mechanism's effective authority cannot contain an operation outside the requesting agent's valid path-specific delegation bound:
	\begin{equation}
		\widehat{E}_i^{\pi}(c;\widehat{\mathbf S},P)\subseteq D_i^{\pi}.
		\label{eq:prop-bound}
	\end{equation}
	\emph{Argument.} Equation~\eqref{eq:mechanism-effective} constructs the effective set by selecting operations \(a\in D_i^{\pi}\) that also satisfy local and upstream restrictions. Thus no policy-state input can add an operation not already contained in \(D_i^{\pi}\). This set-theoretic property does not itself establish that the delegation records are authentic; that remains a trust assumption.
	
	\subsection{Reassessment of Relevant Upstream Changes}\label{subsec:prop-reassessment}
	
	\textbf{Property 2 (conditional downstream restriction).} Fix an authenticated path \(\pi\), a requested operation \(a\), request context \(c\), and policy \(P\). Suppose an accepted security-state record for upstream principal \(u\) changes, with \(u\in R_{\pi}(i,a,c,t)\), while all other accepted decision inputs remain fixed. If the new state causes \(a\notin H_P(u,i,\pi,c,\widehat S'_u)\), the next affected evaluation denies \(a\), even if the downstream credential and \(D_i^{\pi}\) have not changed.
	
	\emph{Argument.} The required conjunction in Equation~\eqref{eq:mechanism-permit} becomes false for principal \(u\). The claim therefore follows for a subsequent mediated request evaluated with the updated accepted record. If an ancestor is not relevant to \(a\), its state change does not by itself alter this particular decision. This is a \emph{policy-scoped} propagation property, not blanket revocation of every descendant.
	
	A stronger restriction-ordering property is available \emph{only if configured by policy}. When \(\widehat S'_u\preceq_{\mathrm{trust}}\widehat S_u\) and the ancestor allowance is monotone, so that
	\begin{equation}
		H_P(u,i,\pi,c,\widehat S'_u)\subseteq
		H_P(u,i,\pi,c,\widehat S_u),
		\label{eq:prop-monotonic-ancestor}
	\end{equation}
	then, with the same path, relevance relation, context, other states, and policy, degrading only \(u\)'s state cannot increase the operations allowed by the mechanism. There is no assertion of monotonicity across changing contexts, different policies, or incomparable state labels.
	
	\subsection{Mediated Decisions, Unavailable State, and Analytical Limits}\label{subsec:prop-boundaries}
	
	\textbf{Property 3 (request-bound enforcement requirement).} Under the assumed non-bypassable PEP, an operation is executed only after the PEP receives a \textsc{Permit} for the corresponding request. The mechanism does not authorize a later request merely because an earlier one was permitted. The property depends on correct PEP/PDP integration; it is not established by the set equations alone.
	
	\textbf{Property 4 (conservative handling of required state).} If a required state input fails the configured provenance or freshness conditions, the proposed default denies the affected operation (Section~\ref{subsec:mech-freshness}). This avoids interpreting a missing or stale required signal as evidence of a normal state. It may also reduce availability when state records are delayed or temporarily unreachable.
	
	These statements distinguish \emph{decision correctness relative to accepted inputs} from \emph{security correctness relative to the actual world}. A false negative from an external monitor, delayed receipt of an incident, corrupted trusted infrastructure, or a tool call outside PEP mediation can invalidate the intended operational protection. Neither the model nor the examples below provide a bound on detection delay or guarantee containment of actions already in progress.
	
	\subsection{Bidirectional Branch Containment}\label{subsec:property-branch}
	
	\textbf{Property 5 (branch-local restriction).} Fix the policy, authenticated delegation graph, and accepted event view. If an event $e$ changes only branch $\gamma$ and the policy restricts its containment allowance from $C_P$ to $C'_P\subseteq C_P$, then the effective authority computed for an affected request on $\gamma$ cannot increase. The effective permission set on a distinct branch $\delta$ remains unchanged provided that $\delta$ is outside $\mathcal{I}_P(e,\pi,a,c)$ and all of its accepted decision inputs remain fixed.
	
	\emph{Argument.} Equation~\eqref{eq:mechanism-effective} intersects the base permission set with the branch allowance. Replacing that allowance by a subset can only preserve or remove operations. The distinct branch uses a separate authenticated context and accepted input snapshot; no assumption of automatic global compromise is needed. The property concerns authorization sets, while execution-level sandbox separation depends on the enforcing environment.
	
	\subsection{Illustrative Delegation-Chain Scenarios}\label{subsec:prop-scenarios}
	
	Consider the authenticated path \(U\!\rightarrow\!A\!\rightarrow\!B\!\rightarrow\!C\), where \(C\)'s delegated bound permits \(\{\mathrm{read},\mathrm{write}\}\) on a resource. Assume normal local checks and a policy making \(A\)'s state relevant to \(\mathrm{write}\) but not \(\mathrm{read}\). Table~\ref{tab:mechanism-trace} (Section~\ref{subsec:mech-trace}) summarizes the resulting decisions for successive requests; its rows are policy outcomes derived from the specified predicates, \emph{not experimental observations}.
	
	The crucial comparison is between two requests by \(C\) whose identity and delegated bound remain unchanged. A local authorization baseline may permit both because it does not consume \(A\)'s changed state, whereas the chain-aware rule denies the affected write after accepting that record. This result follows from the specified policy inputs and does not constitute an empirical attack-success measurement.
	
	\subsection{What the Formal Analysis Does Not Establish}\label{subsec:prop-notproved}
	
	The stated properties are limited to the defined authorization boundary and assumptions. They do not constitute a proof that Agent-ZTA detects upstream compromise, prevents every malicious action, provides immediate global revocation, guarantees availability, or achieves lower attack success in practice. Such claims would require an explicit implementation model and empirical evidence. The analysis instead provides checkable invariants and cases against which any implementation can be assessed.
	
	\section{Limitations and Future Evaluation}\label{sec:limitations}
	
	\subsection{Model Coverage and Evidence Boundaries}\label{subsec:limitations-status}
	
	Agent-ZTA defines authenticated path bounds, accepted principal and branch-state inputs, bidirectional event impact, a deterministic authorization predicate, and branch-specific restricted modes. The set relationships establish bounded authority under the stated policy and trust assumptions. The worked decision traces demonstrate outcomes of those predicates for selected inputs; they are not empirical measurements of attack containment, latency, or sandbox isolation.
	
	The approach relies on established delegation, policy enforcement, and risk-aware authorization foundations. Its distinctive decision structure associates security events with the authenticated dependency branch and uses that association to constrain later authority in either direction. Whether this structure provides an advantage over a well-configured chain-aware baseline requires direct comparison.
	
	\subsection{Assumptions and Practical Constraints}\label{subsec:limitations-assumptions}
	
	The mechanism depends on the correctness and provenance of delegation records, applicable authorization policies, state-to-principal associations, and PEP/PDP enforcement. Agent-ZTA does not produce compromise detections or solve prompt injection by itself: incorrect, incomplete, or delayed external signals can lead to unsafe permits or unnecessary denials. Required-state fail-closed behavior protects against silent optimistic assumptions but can reduce availability. Similarly, reauthorization occurs at mediated decision points; the model does not specify instantaneous state dissemination, cancellation of in-flight tool operations, or recovery of resources affected before a restriction was enforced.
	
	Policy choices remain consequential. Defining which ancestors are relevant to which operations, when a reported condition counts as degraded, and whether recovery restores permissions requires context-specific decisions. Overly broad relevance rules can block legitimate work; overly narrow rules can overlook important upstream dependencies. The present model deliberately does not prescribe a universal risk score, anomaly detector, cryptographic credential format, or distributed coordination protocol.
	
	\subsection{Empirical Evaluation}\label{subsec:limitations-future}
	
	An empirical evaluation can instantiate the specified rules in a controlled multi-agent testbed and compare them with a local/per-request authorization baseline. The comparison should examine downstream restriction after upstream events, parent-workflow reassessment after downstream events, branch isolation behavior, and authorization performance. The rule set and worked cases in this paper define what such an evaluation must measure.
	
	A fair comparison should use the same delegated scopes and tool operations for both mechanisms, vary the source and timing of accepted events, and record changes in authorized operations, unaffected-branch continuity, and request decision cost. The chosen scenario set and testbed implementation are separate from the mathematical results established in this manuscript.
	
	\section{Conclusion}\label{sec:conclusion}
	
	This paper developed Agent-ZTA, a chain-aware Zero Trust authorization and containment model for multi-agent delegation. The model separates original path-specific delegated authority from current effective permissions and binds authorization decisions to accepted security-state events, authenticated workflow dependencies, and deterministic policy rules. It extends upstream-to-downstream reassessment with downstream-to-parent workflow tainting, while preserving branch specificity so that unrelated delegation paths are not restricted solely by association.
	
	The resulting decision procedure incorporates freshness and provenance checks, a fail-closed rule for missing required state, and restricted or jail-mode allowances for affected execution branches. The security analysis states non-escalation, conditional downstream restriction, request-bound enforcement, conservative fail-closed handling, and branch-local restriction properties under the specified authorization predicates. Practical containment additionally depends on correct event inputs, non-bypassable enforcement, and an isolated execution environment wherever physical sandboxing is required. The model therefore provides a complete, checkable specification of bidirectional chain-state authorization, covering its state model, architecture, decision procedure, and security properties.

\end{document}